%% file: ndncert.tex
\documentclass[10pt,sigconf]{acmart}

\usepackage{booktabs}
\usepackage{soul}
\usepackage{xspace}
\usepackage{graphicx}
\usepackage[inline]{enumitem}
\usepackage{ifthen}
\usepackage{etoolbox}
\usepackage{xstring}
\usepackage{tikz}
\usepackage{pifont}
\usepackage{amsmath}
\usepackage{amsfonts}
\usepackage{listings}
\usepackage{color}
\usepackage{tabularx}
\usepackage[normalem]{ulem}

\definecolor{dkgreen}{rgb}{0,0.6,0}
\definecolor{gray}{rgb}{0.5,0.5,0.5}
\definecolor{mauve}{rgb}{0.58,0,0.82}

\include{macros}

\def\cameraReady{} 

\graphicspath{{figures/}}

\begin{document}

\title{On Certificate Management in Named Data Networking}

\ifdefined\cameraReady
\author{Zhiyi Zhang}
\affiliation{
	\institution{UCLA}
}
\email{zhiyi@cs.ucla.edu}

\author{Su Yong Wong}
\affiliation{%
	\institution{UCLA}
}
\email{email@domain.com}

\author{Junxiao Shi}
\affiliation{%
	\institution{NIST}
}
\email{junxiao.shi@nist.gov}

\author{Davide Pesavento}
\affiliation{%
	\institution{NIST}
}
\email{davide.pesavento@nist.gov}

\author{Alexander Afanasyev}
\affiliation{%
	\institution{Florida Int'l University}
}
\email{aa@cs.fiu.edu}

\author{Lixia Zhang}
\affiliation{%
	\institution{UCLA}
}
\email{lixia@cs.ucla.edu}
\else
\author{Anonymous}
\fi

\renewcommand{\shortauthors}{Anonymous}

\input{sections/abstract}

\maketitle

\input{sections/intro}
\input{sections/background}

\input{sections/ndn-sec}

\input{sections/concepts}

\input{sections/model}

\input{sections/overview}
\input{sections/evaluation}
\input{sections/challenge-discussion}

\input{sections/conclusion}

\bibliographystyle{acm}
\bibliography{ndncert}

\end{document}

%% file: macros.tex
\newcommand{\todo}[1]{\mynote{TODO:}{#1}{blue}}

\newcommand{\zz}[1]{\mynote{Zhiyi}{#1}{magenta}}
\newcommand{\cAA}[1]{\mynote{Alex}{#1}{orange}}

\usepackage{ifthen}

\newboolean{showcomments}
\setboolean{showcomments}{true}
\ifthenelse{\boolean{showcomments}}
{ \newcommand{\mynote}[3]{
    \protect\fbox{\bfseries\sffamily\scriptsize#1}
    {\small$\blacktriangleright$\textsf{\emph{\color{#3}{#2}}}$\blacktriangleleft$}}}
{ \newcommand{\mynote}[3]{}}

\newcommand\sysname{NDNCERT\xspace}

\newcommand\data{Data\xspace}

\newcommand\para[1]{\vspace{0.05in} \noindent \textbf{#1.}}

\def\first{({i})\xspace}
\def\second{({ii})\xspace}
\def\third{({iii})\xspace}
\def\fourth{({iv})\xspace}

\definecolor{verylightgray}{gray}{0.8}

\usepackage{etoolbox}
\usepackage{xstring}
\DeclareListParser{\doslashlist}{/}
\newcounter{ndnNameComponentCounter}%
\newcommand{\name}[1]{{%
  \setcounter{ndnNameComponentCounter}{0}%
  \renewcommand{\do}[1]{{%
    \ifnumgreater{\value{ndnNameComponentCounter}}{0}{\allowbreak/}{}%
    \ifnumodd{\value{ndnNameComponentCounter}}{}{}%
    \detokenize{##1}}%
    \stepcounter{ndnNameComponentCounter}}%
``{\fontfamily{cmtt}\small\selectfont\IfBeginWith{#1}{/}{/}{}\doslashlist{#1}}''%
}}

\usepackage{etoolbox}
\usepackage{xstring}
\newcommand{\namesm}[1]{{%
  \setcounter{ndnNameComponentCounter}{0}%
  \renewcommand{\do}[1]{{%
    \ifnumgreater{\value{ndnNameComponentCounter}}{0}{\allowbreak/}{}%
    \ifnumodd{\value{ndnNameComponentCounter}}{}{}%
    \detokenize{##1}}%
    \stepcounter{ndnNameComponentCounter}}%
``{\fontfamily{cmtt}\tiny\selectfont\IfBeginWith{#1}{/}{/}{}\doslashlist{#1}}''%
}}

%% file: sections/abstract.tex
\begin{abstract}


Named Data Networking (NDN) secures network communications by requiring all data packets to be signed when produced.
%
This requirement necessitates efficient and usable mechanisms to handle NDN certificate issuance and revocation, making these supporting mechanisms essential for NDN operations.
In this paper, we first investigate and clarify core concepts related to NDN certificates and security design in general, and then present the model of NDN certificate management and its desired properties.
We proceed with the design of a specific realization of NDN's certificate management, NDNCERT, evaluate it using a formal security analysis, and discuss the challenges in designing, implementing, and deploying the system, to share our experiences with other NDN security protocol development efforts.

\end{abstract}

%% file: sections/intro.tex
\section{Introduction}

In public-key cryptography, public key certificates assert the binding between a public key and a communicating identity. As such, certificate management for issuance, storage, delivery, renewal, and revocation, is a critical task.
%
%
Named Data Networking (NDN)~\cite{ndn2014} architecture embeds the use of the public-key cryptography into its design to ensure the integrity and authenticity of the communication.
Specifically, all data packets are named and signed at the time of production, securely binding the name and the content of the packet.
Following the security principle of ``least privilege''~\cite{saltzer1975protection}, NDN promotes the use of hierarchies of signing keys, each with limited-scope (e.g., \name{/foo/bar1/KEY/..}, \name{/foo/bar2/KEY/..}, ... keys to operate only within \name{/foo/bar1} and \name{/foo/bar2} namespace respectively) to minimize potential damage in case any private keys are compromised.
Therefore, \emph{a usable, flexible, and scalable certificate management} is among the foremost important steps in realizing NDN security and NDN architecture in general.

In this paper,
we clarify several core concepts that are tightly related to the design of NDN security protocols (\S\ref{sec:clarifications}), which are often misunderstood/misinterpreted (even among ourselves) due to the influence by today's security practices (\S\ref{sec:motivation}).
We then provide a systematic study of the new requirements and the challenges for certificate management in NDN (\S\ref{sec:model}).

Guided by the identified requirements, we propose \sysname, an automated certificate management system for certificate issuance and revocation in NDN (\S\ref{sec:design}).
To ensure the security of the protocol and the usability of the implementation on different hardware platforms, we performed a basic formal analysis of the proposed protocol and overhead study of our prototype implementations.
%
%
We end the paper by sharing identified challenges and lessons learned from realizing and deploying an earlier version of \sysname~\cite{zhang2017ndncert}, which has been running over the NDN Testbed for more than a year.

The contributions of our work are threefold:
\first a clarification of several core concepts in designing NDN security protocols;
\second a systematic study of the new requirements and the challenges of certificate management in NDN; and
\third the design and development of \sysname.
In addition to a specific protocol design, we share our experiences in addressing the challenges we met in realizing and deploying the system.
We hope our experiences can be of more value contributing to the research and development of security solutions in general, and to the NDN rollout in particular.


%% file: sections/background.tex
\section{The Use and Management of Certificates in Today's Internet}
\label{sec:motivation}

Certificate issuance and management are essential in any systems that deploy cryptographic protection.
The Internet today relies on TLS to secure communications.
Other application systems such as SSH~\cite{ssh} and GPG~\cite{pgp, gpg} also adopted the use of digital certificates.
Below we use TLS certificate system as a representative of today's widely deployed certificate management system.

Until recently, most TLS certificates have been issued by a relatively small number of commercial Certificate Authorities (CAs)~\cite{aas2019let}.
Commercial CA services came into existence in the mid-90s to meet the need of cryptographic protection for the emerging e-commerce applications at the time.
Pragmatic solutions were also developed to install the self-signed certificates of these CAs into end-users' devices as \emph{trust anchors}, largely through side channels (e.g., web browser installation and updates), and behind the back of users.
Rapid growth for certificate services also leads to the rise of hierarchical relationships between CA providers, for example with CA-1 issues a certificate to CA-2 which in turn issues certificates to end-users.
By and large one could view CAs in the TLS certificate management as trust anchors\footnote{In practice, a trust anchor can run multiple CAs whose certificates are derived from the trust anchor certificate.}, whose certificates get installed into user devices.


The research community has identified several longstanding issues with today's CA practice, which we summarize below.

\para{Authenticity vs. Trust}
There are different views regarding the meaning of a certificate $C_P$ issued to party $P$ by a CA.
Most commercial CA providers hold the view that if $P$ holds a certificate $C_P$ issued by them (especially if it is an Extended Validation/EV certificate), then $C_P$ certifies \emph{both} $P$'s identify and its trustworthiness.
Another view, shared by LetsEncrypt, today's de facto largest CA issuing free-of-charge certificates using automated means~\cite{letsencrypt,acme}, believes that a valid TLS certificate only certifies the authenticity of party $P$'s identity, but not $P$'s trustworthiness.
For example, a phishing website can successfully obtain a valid TLS certificate from a CA, commercial, or otherwise, as long as it owns a domain name and has an available web server~\cite{abuseletsencrypt}.


\para{Externality and Trust}
Today's CAs are \emph{external} and thus agnostic to the communicating parties who need to establish trustworthy relations~\cite{smetters2009securing}.
Anyone can get a certificate, $n$ unrelated parties $P_1$, $P_2$ $\cdots$ $P_n$ possessing a certificate each is unrelated to whether they trust each other.
The fact that the trust anchors of these certificates are decided by yet another \emph{external} entity (e.g., browser vendors) behind the back of end users further argues that certificates serve the purpose of authentication at best, not trust relations.

\para{Constraints}
In today's practice, a certificate is issued to a site (a company, or at least a DNS name owner), and the certificate owner is not allowed to issue certificates for its own sub-namespaces, making it difficult to support the the principle of ``minimal privilege".
Another practice is setting a prolonged certificate lifetime: generally a few months as a minimum, and often one year or even longer--this is especially true in case of manually processed certificates.
Compounded with the coarse certificate granularity, a long lifetime reduces certificates' resiliency against brute force usage-analysis attacks.

%% file: sections/ndn-sec.tex
\section{Names and Certificates in NDN}
\label{sec:background}
NDN~\cite{ndn2014} changes the Internet communication model from IP's pushing packets to destination addresses to fetching data by names.
It also promotes the use of application-layer namespace in network communication, which is a sharp departure from today's protocol stack design, where each layer defines its own identifier space (e.g. IP uses IP addresses, transport protocols use port numbers and sequence numbers).
Naming data with application-layer namespace enables NDN to utilize the semantically meaningful names to simplify and streamline the use of cryptographic data protection as explained in~\cite{zhang2018security}.

\begin{figure}[htbp]
	\centering
	\includegraphics[width=0.4\textwidth]{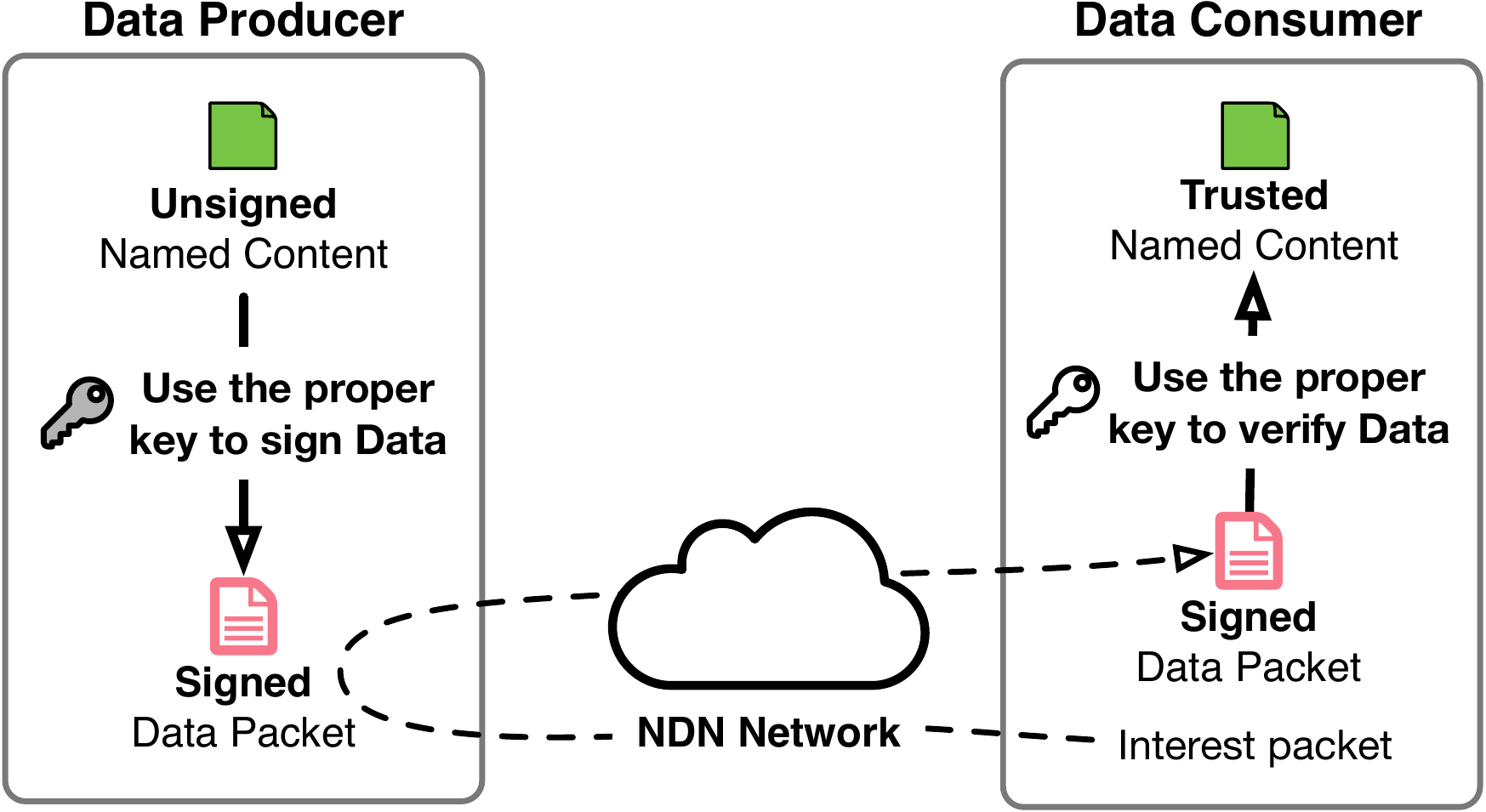}
	\caption{Signature Verification in NDN}
	\label{fig:ndn-signature}
\end{figure}

To secure data directly, producers cryptographically sign data packets, binding their names with the payload (Figure~\ref{fig:ndn-signature}). When a consumer $C$ fetches a \data packet $P_D$, it verifies the signature in $P_D$ to check its authenticity and integrity.
$C$ uses the \emph{key locator}, which contains signing key information, carried in $P_D$ to retrieve the corresponding public key certificate.
$C$ considers $P_D$ trustworthy only if the whole certificate chain is valid and anchored at a certificate specified by $C$'s trust policies~\cite{trust-schema}.

%

\subsection{Names and Configurations}
%

An NDN entity requires four pieces of information to become an active participant in an NDN environment, i.e., being able to publish and fetch data~\cite{zhang2018security}.
They are
\first the name assigned to the entity,
\second a certificate that binds the entity's name and its public key,
\third a trust anchor which is represented as a certificate, and
\fourth trust policies~\cite{trust-schema}.

While pure consumers only need \third and \fourth to participate, entities in general are most likely to act in both capacities (e.g., at least being a producer for local management tasks), therefore all four pieces should be considered essential.
Furthermore, NDN utilizes the fact that everything in the cyberspace can be encoded as a named piece of data\footnote{Here we use ``naming'' broadly to cover various identifiers.}.
Therefore certificates and trust policies can simply be encoded as named, secured data packets that can be fetched as any other content.
%

\subsection{NDN Certificate}
\label{sec:ndn-certificate}


An NDN certificate binds an entity's name and its public key. It is a piece of \data signed by the issuer, and its name follows the naming convention below.
\begin{center}
	/<identity-name>/KEY/[key-id]/[info]/[version]\footnote{<> represents one or more name components while [] refer to one name component.}
\end{center}
For example, if a user Alice is assigned a name \name{/example/alice} and has a public key \name{/example/alice/KEY/123/}, a corresponding certificate issued by the issuer ca-1 should be \name{/example/alice/KEY/123/ca-1/456}. This certificate allows Alice to produce verifiable Data packets using the certified key. Once consumers fetch her data, they can authenticate Alice's public key through the certificate and verifies data. 
Alice can also further assign sub-namespaces (e.g. /example/alice/guest/) to other entities and issue corresponding certificates signed by her certified key.

One can easily define various trust relations between certificates because of the semantic meanings embedded in their names.
For example, adding a component \name{guest} to an entity's name $E_g$, one can easily define trust policies to make $E_g$'s certificate represent a lower privilege level, compared to other certificates with a component \name{member}. 
In this way, one can define policies based on the certificates' names to express different levels of trustworthiness~\cite{trust-schema}. 
As another example in vehicular networking, geolocation information can be embedded into the name of a certificate issued to a vehicle passing a specific location
as follows: \name{/vehicles/geo: 34n-118w/id:123/KEY/456/ca-1/789}, certifying that a vehicle 123 physically appeared at the location.

\subsection{Certificate Validation}

Cryptographic verification of the data signature requires the whole certificate chain from the signer's certificate to the trust anchor certificate specified in the trust schema. To pass the verification, all the involved certificates must satisfy the following requirements.
\begin{enumerate} [leftmargin=*]
	\item All certificates in the certificate chain must be available. In other words, the verifier must be able to fetch them if they are not already cached locally.

	\item All certificates in the chain must be valid, i.e. neither expired nor revoked.
\end{enumerate}
Once these two requirements are met, the integrity of the data is verified. 
\S\ref{sec:design:availability} and \S\ref{sec:design:main} will describe in detail the certificate availability and revocation mechanism provided by \sysname.

%% file: sections/concepts.tex
\section{Relations between Names, Certificate, and Trust}
\label{sec:clarifications}

Before moving onto the NDN certificate management design, we would like to first clarify several core concepts that relate to each other:
\begin{enumerate*}
	\item the relation between name assignments and certificate management;
	\item the relation between a certificate holder and a certificate issuer;
	\item the relation between a certificate issuer and a trust anchor; and
	\item the relation between authentication and trust.
\end{enumerate*}


\subsection{Name Assignment and Certificate Management}
\label{sec:name-cert}
As we discussed in \S\ref{sec:background}, NDN utilizes application-layer names in both network layer communication and security protection solutions.
Name assignment needs to be done out-of-band due to two reasons:
\first~assigning a name to an entity $E$ requires the authentication of $E$, but $E$ does not have an identity yet; and
\second~names are generally semantically meaningful and thus name assignments need to be done by entities who understand the semantics.

It is possible to issue a certificate to a named entity $E$ at the time of $E$'s name assignment.
However, certificate issuance requires in-band operations: $E$ must be running, generate its own key pairs, and communicate with the issuer to get a certificate.
This creates challenges if one has to combine this in-band step with the out-of-band name assignment.

Compared with name assignment, the certificate issuance can be done online through a standard process (e.g., \sysname proposed in this paper) once $E$ obtains its name and trust anchor by some out-of-band means, and certificates also require continued management after issuance.

As such, this paper focuses on the design of a usable NDN certificate management protocol, leaving the name management to the namespace owner.
In the implementation of \sysname, we also provide mechanisms to facilitate name assignments whenever feasible in different scenarios as discussed in \S\ref{sec:discussion:iot}.


\subsection{Allowing A Certificate Holder to Issue Certificates}
\label{sec:issue-cert}

NDN utilizes application-layer's structured, semantically meaningful names to allow users and applications to define a rich set of trust policies, and to support least privilege principles in cryptographic key usage.
Therefore, it is important to allow a certificate holder to issue derived certificates for the sub-namespace under its own name.
Such structured delegation provides a foundation for flexible privilege separation and limiting the scope of individual keys, based on name relationship among entities captured by the trust policies (e.g., trust schema~\cite{trust-schema})



\begin{figure}[t]
	\centering
	\includegraphics[width=0.25\textwidth]{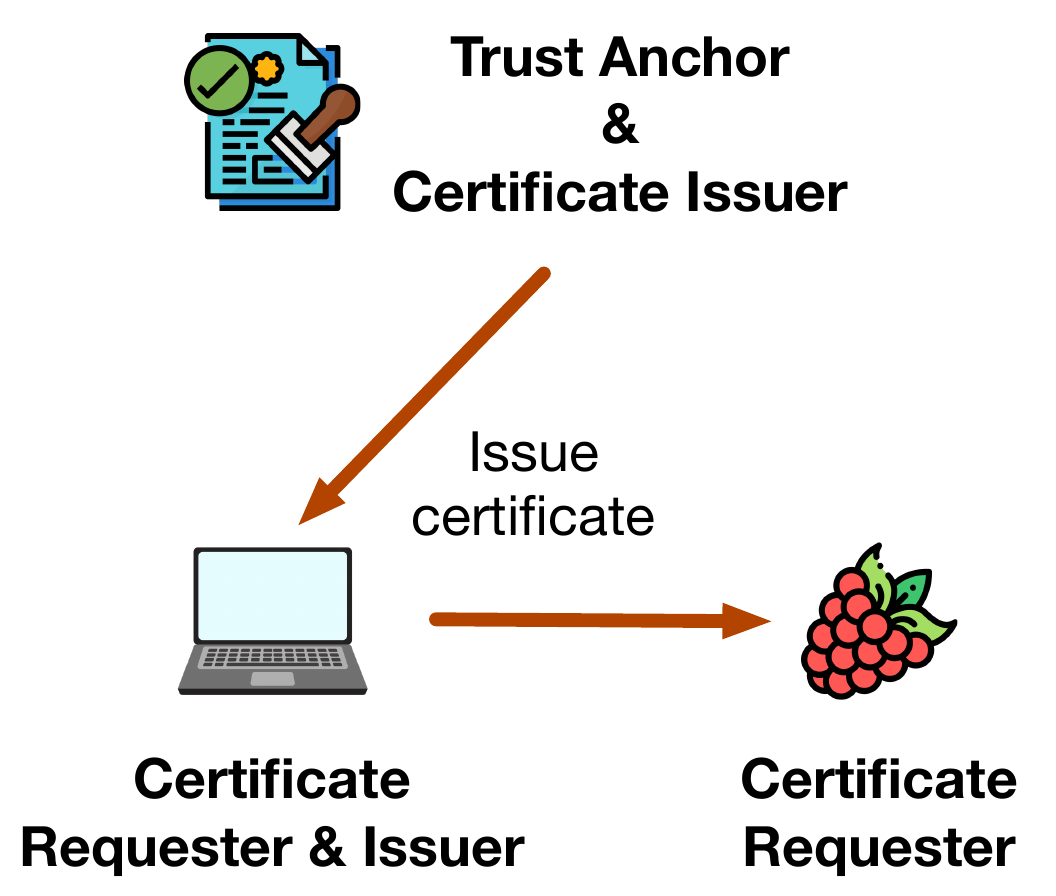}
	\caption{The laptop issues a certificate to the IoT device as a certificate issuer rather than a trust anchor}
	\label{fig:trust-anchor-issuer}
\end{figure}

We use an example to illustrate the advantage from allowing a certificate holder to issue certificates: assume a professor Alice is issued a certificate by the campus.
She wants to let her new IoT device (e.g., a Raspberry Pi) produce some experimental data for her students.
Assuming all entities on campus are configured with the campus key as the trust anchor, to generate data that can be verified by her students, Alice could either
\first apply for another certificate for her device directly from the campus certificate issuer ,or
\second copy the existing certificate and private key from her laptop to the device.
The first approach leads to scaling concerns: the campus key would be used to issue too many certificates, and when the key changes, it may be impossible to re-issue new certificates to all the devices on campus.
The second one introduces great security risks for Alice's laptop.
Allowing Alice to derive another certificate directly from her existing certificate (Figure~\ref{fig:trust-anchor-issuer}) provides a simple and scalable solution, supports the least privilege security principle.

\subsection{Certificate Issuer $\neq$ Trust Anchor}

A trust anchor represents an administrative domain (e.g., a home, a campus) and its certificate must be configured, generally via a secure out-of-band means, into all entities within the domain.
If a campus certificate \name{/campus1/KEY/self-signed/123} is configured as a trust anchor certificate, an issuer's certificates like \name{/campus1/alice/KEY/...} can be trusted by other users on the campus if:
\first~their certificate chain terminates at \name{/campus1/KEY/self-signed/123}, and
\second~their trust policies consider such chain valid with respect to signer-signee name relations.
Although Alice can serve as a certificate issuer for her IoT devices (Figure~\ref{fig:trust-anchor-issuer}), Alice's certificate is not the ``root of trust'' on the campus.

Generally speaking, although a certificate issuer must be trusted by its requesters, the issuer may, or may not, be a trust anchor in the system.
Recall that trust anchors need to be configured via secure out-of-band means and changes to trust anchors can be expensive to handle; therefore, one must be careful in deciding trust anchors.

\subsection{Authentication $\neq$ Trust}
\label{sec:clarifications:trust}
Authentication is required in establishing trust relations, but it does not equal to trust.
Authentication confirms the identity of a party while trust decides what actions an authenticated party is trusted to perform.
NDN security design develops \emph{separate} mechanisms to support authentication and trust:
certificate verification checks the authenticity of each entity, and
trust schemas are used to determine the allowed actions for each authenticated entity, as explained in~\cite{trust-schema,zhang2018security}.

For example, a smart home system can issue different certificates for a home member and a guest. Its trust policy would allow only a home member to unlock the smart door while allowing both the member and the guest to control the house light.
In this case, even though both parties are authenticated, the trust on each is separately determined by one's identity and policy.

%% file: sections/model.tex
\section{NDN Certificate Management Model}
\label{sec:model}

In this section, we formally model the certificate management system in NDN, summarize the assumptions, and define the desired properties of such a system.

\subsection{System Model}

A certificate management procedure involves three main logically separated parties: the certificate requester, the certificate issuer, and the name authority.

\begin{itemize} [leftmargin=*]
	\item \textbf{Certificate Requester}:
	an entity that requests a certificate to prove its ownership of a namespace and bind its public key to the namespace.

	\item \textbf{Certificate Issuer}:
	an entity that validates and processes certificate issuance/renewal/revocation requests under the associated namespace.

	\item \textbf{Name Authority}:
	an entity that manages an NDN namespace and determine whether a requester is an owner or is allowed to own a particular sub namespace.
\end{itemize}

\subsection{Assumptions}

Our model of the NDN certificate management makes the following  assumptions:
\begin{enumerate} [leftmargin=*]
\item A certificate issuer is available for the corresponding certificate requesters. Note that not all issuers are assumed to be available to all requesters.

\item The trust anchor is pre-installed by an out-of-band way, and therefore a requester can authenticate an issuer.

\item There is a secure channel between the name authority and the certificate issuer, and thus they can cooperate to verify the legitimacy of requests. For example, the certificate issuer can access a name authority's database of short term secure codes.
\end{enumerate}

\subsection{Desired Properties}

We define the desired properties of a certificate management system in NDN as follows.

\para{Usability}
NDN's certificate management system should provide an easy to use application for certificate issuance, renewal, and revocation. Specifically, its design should minimize the necessity of manual operations and configurations.

\para{Flexibility}
The certificate management should be adaptable to different application scenarios because
\first each system can have different identity verification means;
\second secure channel between certificate issuers and name authorities can be realized in various forms;
\third name authorities can have diverse certificate naming conventions.

\para{System Security}
As an essential part of network security, the certificate management process itself should be secured.
\begin{itemize} [leftmargin=*]
	\item Communication security.
	The system must ensure authenticity, integrity, and confidentiality of communication between an issuer and a requester; in addition, forward secrecy, resistance to replay attacks and man-in-the-middle attacks should also be provided.

	\item Issuer security.
	While an issuer should maintain its availability to legitimate requesters, its exposure to potential compromise must be minimized, e.g., by security hardware~\cite{sgx, trust-zone} or multi-factor authentication (MFA) to protect an important issuer private key.
	Note that issuer security will not be covered by the design proposed in this paper and is left for specific applications.
\end{itemize}

%
%
%
%
%
%
%

\para{Availability}
Issued certificates must be highly available to data consumers who need to verify \data packets.
If there exists any revoked certificates, the availability of revocation records must be realized as well.

\para{Scalability}
The certificate management system must be scalable to process certificate operation requests in large and complicated systems.

\para{Transparency}
Since issuer's correctness directly impacts the security of the whole system, the certificate issuance, renewal, and revocation should generate immutable and publicly auditable logs~\cite{cert-transparency}.

%% file: sections/overview.tex
\section{Design of \sysname}
\label{sec:design}

\begin{figure}[t]
	\centering
	\includegraphics[width=0.45\textwidth]{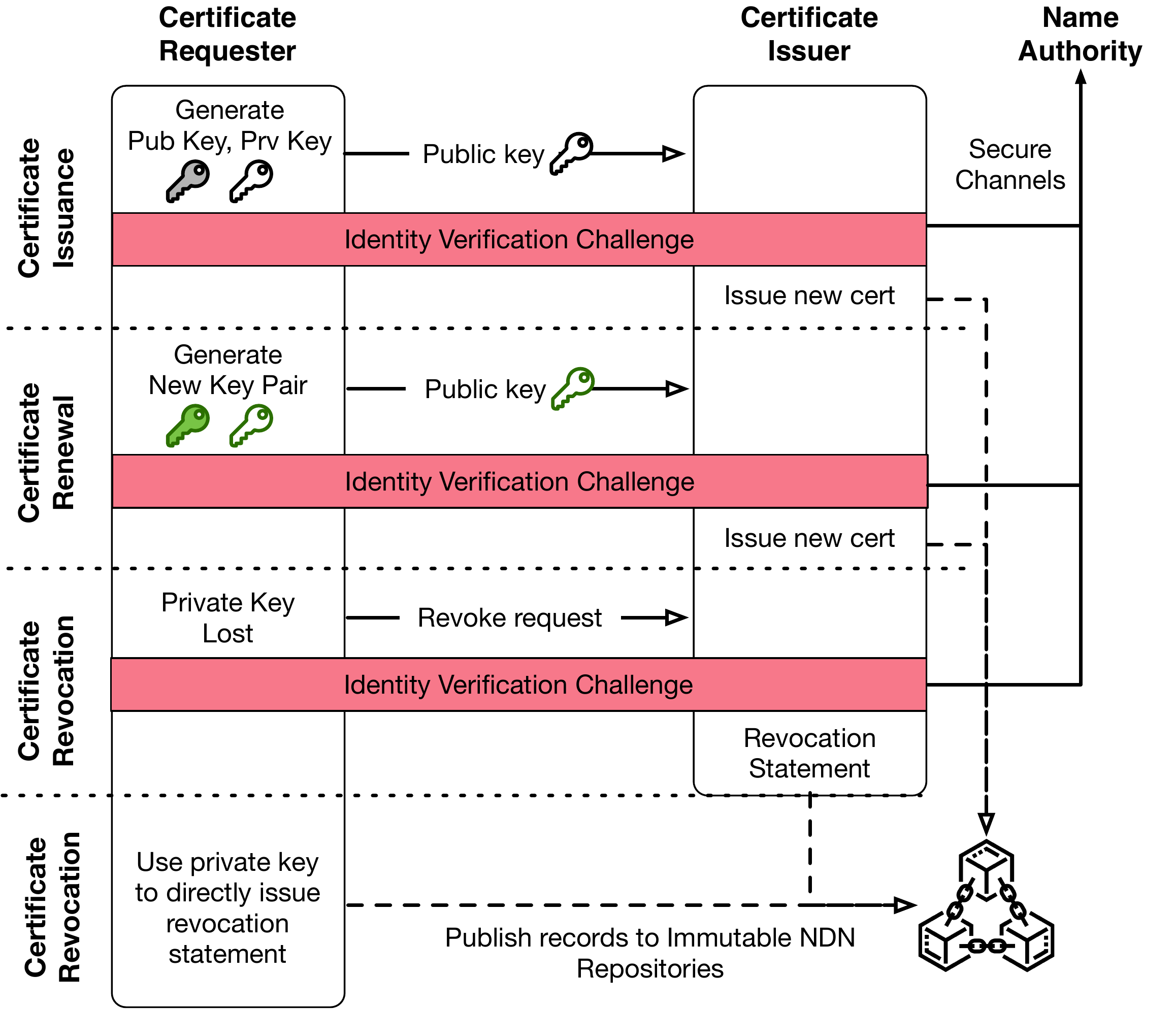}
	\caption{An Overview of \sysname}
	\label{fig:ndncert-logic}
\end{figure}

Based on the system model and desired properties, we propose the design of \sysname (Figure~\ref{fig:ndncert-logic}) to allow ease-of-use certificate issuance, renewal, and revocation.
In addition, we also present how flexibility, transparency and certificate availability can be realized, and \sysname's security considerations.

In \sysname, a certificate requester communicates with a certificate issuer over a public (insecure) network for certificate issuance, renewal, and revocation.
The certificate issuer can also contact the name authority through a pre-established secured channel in either in-bind or out-of-band manner.

In addition to a requester, issuer, and name authority, \sysname also assumes the existence of NDN repositories (NDN repo) that support immutable logging. To provide immutable logging, the distributed ledger technologies (DLT) or data storage system with strict access control can be executed over NDN repos. We do not assume a specific implementation of the immutable logging because it is out of the paper's scope.

\subsection{Certificate Issuance, Renewal, and Revocation}
\label{sec:design:main}

\sysname provides a usable means to issue, renew, and revoke a certificate.

\para{Certificate Issuance}
In the certificate issuance procedure, the issuer verifies the following information from the requesters, assuming that a requester has already obtained a name for the certificate request from the name authority.

\begin{enumerate} [leftmargin=*]
	\item The ownership of the private key whose public key will be certified by the issuer.
	\item The legitimate identity to own a certificate of a requested name.
\end{enumerate}

The ownership of the private key is verified by asking the requester to use the private key to sign the request with a nonce or timestamp (to prevent replay attacks).
Then, the requester's identity information is checked through an \emph{identity verification challenge} and with the help of the name authority.
\sysname supports both out-of-band and in-band challenges.
For example, for an out-of-band challenge, the requester can physically meet and obtain the secret from the name authority, while for an in-band case, the requester can prove its identity by showing the already owned certificate from another issuer trusted by the current issuer.

After the requester proves both ownership of the private key and the legitimate identity, the issuer will sign the certificate for the requester.
The certificate can be exported to NDN repositories for high availability and the issuance record can optionally be appended to immutable logging systems for auditing purposes.

Certificate renewal can be considered to be a special case where the identity verification challenges require an ownership proof of a previous key; thus, in principle, \sysname does not separate certificate renewal from issuance.
Since the certificate renewal process can be automated with \sysname, using short-lived certificates is recommended.

\para{Certificate revocation}
As mentioned, \sysname recommends using and renewing short-lived certificates to minimize the use of explicit certificate revocations, thus reducing system complexity and status checking overhead for issuers, requesters, and consumers. At a high-level, the idea resembles OCSP stapling~\cite{ocsp}; however, instead of using a special OCSP response, \sysname simplifies the process by letting issuers directly issue short-lived certificates. When a certificate is supposed to be revoked, the issuer can simply stop renewing the certificate.

When desired, \sysname can realize an explicit certificate revocation.
Following the security practices, the following three types of entities should be able to issue a revocation:
\begin{enumerate}[leftmargin=*,noitemsep,topsep=1pt,parsep=0pt,partopsep=0pt]
	 \item The issuer of the certificate.

	 \item The owner of the namespace.
	 In normal cases, this refers to the certificate or the corresponding private key holder; however, when the private key is compromised, the owner of the namespace can prove its identity to the issuer and ask the issuer to generate a revocation.

	 \item The holder of the corresponding private key.
	 Normally, this refers to the namespace owner; when the key is compromised, the attacker can also access the private key.
	 As such, the private key holder can directly use the private key to sign a revocation.

\end{enumerate}
A revocation, which is also a Data packet, should be signed by the private key of either the certificate to be revoked or its issuer.
Importantly, the revocation message should be appended to the immutable logging system available to the whole system.

%
%

\subsection{Flexibility}
\label{sec:design:configurability}

In order to allow different systems to configure \sysname for their own application scenario, \sysname makes the identity verification a separate, extendable module, allowing issuer operator to add/remove/update verification challenge means in a plug-and-play manner.

To be more specific, \sysname provides a general framework to carry on different identity verification challenges and defines a set of unified interfaces that allow developers to develop new challenges.
In addition, we also provided several default identity verification challenges, including proof of the ownership of a PIN code, an email address, a private key, and an existing certificate issued by another issuer.

When the naming assignment policy can be represented by a set of rules, \sysname also allow the name authority to pass these rules into the issuer via a callback function, simplifying the communication between the name authority and the certificate issuer.

\subsection{Scalability}

In \sysname, the preferred mechanism to improve scalability from a system perspective is hierarchical delegation of certificate management.
To be more specific, instead of letting a single centralized issuer manage all the certificates for the entire system, multiple issuers under the root namespace can be established.
This not only brings better scalability in terms of each issuer's overhead and management complexity, but also allows a finer granularity of name assignment, privilege separation, and trust relationship as discussed in \S\ref{sec:issue-cert}.
In \S\ref{sec:deployment}, our trial deployment on the NDN testbed shows a live example of such delegation.

\subsection{Transparency}
\label{sec:design:immutable}

Transparency through immutable logging is crucial for the certificate management system as it provides accountability and auditability (e.g., to detect intrusion), and improves availability of important records.

In \sysname, certificates issuance, renewal, and revocation records are immutably recorded.
Immutability can be realized in several ways, e.g., immutable database, distributed ledger technologies (DLT).
For example, following the concept of DLedger~\cite{dledger}, records can be linked together with their names and hash value; when multiple parties are involved, records generated from different parties can interlock each other for better security.

%
%

\subsection{Certificate Storage and Availability}
\label{sec:design:availability}

\sysname does not specify how certificate availability should be realized. We present several mechanisms that can be applied.

\begin{itemize} [leftmargin=*]

	\item Holder-side certificate availability:
	Each certificate holder should serve their own certificates. In addition, when the certificate holder is also a producer, it can also provide the data and certificates~\cite{certificatebundle} together to ensure certificate availability.

	\item NDN repo:
	NDN repo, dedicated network-layer storage, can improve the availability of NDN certificates, e.g., when certificate holder are offline.
	For example, a certificate issuer can export the certificate data packet to an NDN repo when it issues a certificate.

      \item In-network cache:
	Since a certificate is an NDN data packet, it can be cached at network routers.
	In addition to the passive network cache~\cite{nfd}, new cache policies can be developed to improve the cache priority of certificate Data packets.

	\item Immutable logging:
	The immutable logging system can also be used to keep records for original certificate Data packets.
\end{itemize}

\subsection{Security Considerations}
\label{sec:design:security}

To secure the communication between the requester and the issuer, \sysname implements the following mechanisms. First, to ensure the authenticity and integrity of the communication, every Interest packets from the requester are signed by the private key of the requested certificate, and every Data packets from the issuer are signed with the issuer's private key.

Second, to achieve confidentiality and forward secrecy of the communication, the requester and the issuer will jointly create a shared secret key (with Elliptic Curve Diffie Hellman (ECDH)~\cite{ecc} and HMAC-based Key Derivation Function (HKDF)~\cite{hkdf}) to encrypt the subsequent communications (with AES GCM~\cite{gcm} encryption scheme).

%% file: sections/evaluation.tex
\section{Implementation, Evaluation, and Trial Deployment}
\label{sec:impl-evaluation}

\subsection{Implementation}

We have implemented \sysname{}\footnote{The codebase and protocol specification will be published in the final version of the paper} in C++ for general purposes and TypeScript for modern web scenarios.
We provide both the library for developers to integrate \sysname protocols into their specific applications and also command-line tools that can directly be used.
Our implementation supports storage for request states using both memory and persistent storage (e.g., SQLite database).
It can also seamlessly work with existing NDN repo implementations~\cite{python-repo} to improve certificate availability.

As specified in \cite{ndncertspec}, the core \sysname protocol is mainly implemented into two main steps: \textsf{NEW} step and \textsf{CHALLENGE} step.
Table~\ref{table:notations} and table~\ref{table:model} give an symbolic representation of these two steps.

\begin{table}[t]
	\footnotesize
	\begin{tabular}{>{\centering\arraybackslash} m{1.5cm} | >{\centering\arraybackslash} m{6cm}}
		\hline
		\textbf{Notation} & \textbf{Explanation} \\ \hline \hline
		R & Certificate requester \\ \hline
		I & Certificate issuer \\ \hline
		$Nonce$ & Randomly picked nonce in Interest packets. \\ \hline
		$PubKey$ & A public key. \\ \hline
		$PrvKey$ & A private key. \\ \hline
		$g^x, g^y$ & $x, y$ are random numbers and $g$ is a generator in an elliptic curve group for Diffie Hellman protocol. \\ \hline
		$RequestID$ & A random identifier selected by the issuer to identify the session. \\ \hline
		$SIGN(d, k)$ & The digital signature signed by key $k$ over the data $d$. \\ \hline
		$ENC(d, k)$ & The ciphertext of $d$ encrypted with key $k$. \\ \hline
		$KDF(d)$ & The key derived from $d$ with a cryptographically secure key derivation function. \\ \hline
		$Secret$ & Sensitive data exchanged in the identity verification challenge. \\ \hline
	\end{tabular}
	\smallskip
	\caption{Notations}
	\vspace{-0.5cm}
	\label{table:notations}
\end{table}

\begin{table}[t]
	\footnotesize
	\begin{tabular}{>{\centering\arraybackslash} m{1.5cm} | >{\centering\arraybackslash} m{6cm}}
		\hline
		\textbf{Packet Exchange} & \textbf{Modeled Payload\newline(including information carried in the Name)}  \\ \hline \hline

		R $\rightarrow$ I, \namesm{NEW} Interest & $Nonce_1$, $PubKey_{R}$, $g^x$ SIGN(($Nonce_1, PubKey_{R}, g^x$), $PrvKey_{R}$)  \\ \hline

		R $\leftarrow$ I, \namesm{NEW} Data & $Nonce_1$, $RequestID$, $g^y$ SIGN(($Nonce_1, RequestID, g^y$), $PrvKey_{I}$)  \\ \hline

		R $\rightarrow$ I, \namesm{CHALLENGE} Interest & $Nonce_2$, $RequestID$,
		\newline $C_1$ = ENC($Secret_1$, KDF($g^{xy}$)),
		\newline SIGN(($Nonce_2, RequestID, C_1$), $PrvKey_{R}$) \\ \hline

		R $\leftarrow$ I, \namesm{CHALLENGE} Data  & $Nonce_2$, $RequestID$,
		\newline $C_2$ = ENC($Secret_2$, KDF($g^{xy}$)),
		\newline SIGN(($Nonce_2, RequestID, C_2$), $PrvKey_{I}$) \\ \hline
	\end{tabular}
	\smallskip
	\caption{Modeled \sysname Protocol in CPSA}
	\vspace{-0.5cm}
	\label{table:model}
\end{table}

\para{NEW step}
The requester first sends an Interest packet carrying the requested namespace, the requested validity period of the certificate, the public key to be certified, and a signature generated by the corresponding private key.
In reply, the issuer sends a Data packet carrying $RequestID$ which is randomly picked to identify the request instance and a list of identity verification challenges for the requester to pick.
Note that both Interest and Data also contain the public information used for the ECDH key agreement and are signed with the timestamp and nonce to prevent replay attack.

\para{CHALLENGE step}
There can be multiple round trips for the requester to finish the challenge.
In each round trip, the requester sends an Interest packet carrying the selected challenge and the information used in the challenge, and the issuer replies with further requirements or status information specified by the selected challenge.
All the parameters carried by the Interest and Data will be encrypted using the key negotiated at the end of NEW step.
Similar as in NEW step, all the packets are signed by the sender and verified by the receiver.

Besides NEW and CHALLENGE, our implementation also supports other optional steps and supporting features.
For example, we use Realtime Data Retrieval (RDR) protocol~\cite{rdr} for requesters to discovery new versions of issuer's profile and download it.
In addition, \sysname allows redirection, which allows an issuer to redirect its requesters to trusted sub namespace issuers, providing better scalability and flexibility.

\subsection{Evaluation}

\begin{figure}[t]
	\centering
	\includegraphics[width=0.46\textwidth]{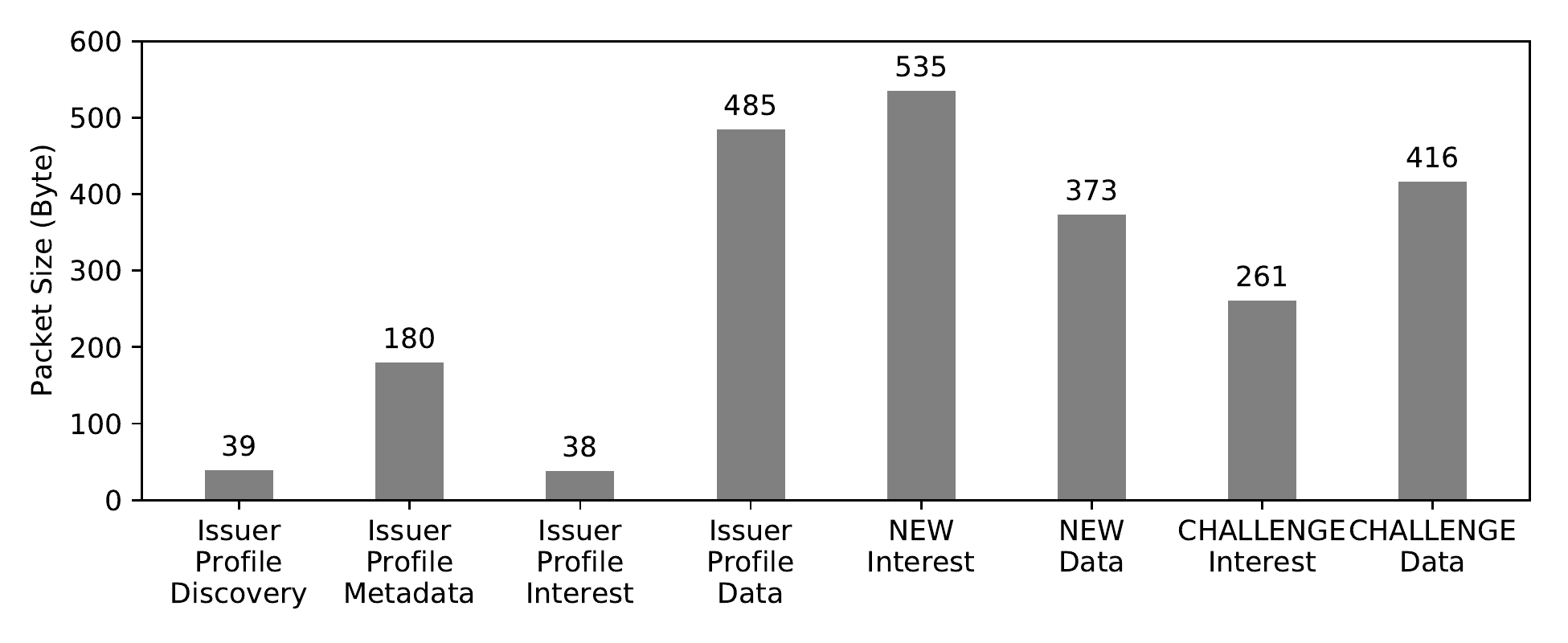}
	\caption{Packet Size Measurement}
	\vspace{-0.3cm}
	\label{fig:pkt-size}
\end{figure}

\begin{figure}[t]
	\centering
	\includegraphics[width=0.46\textwidth]{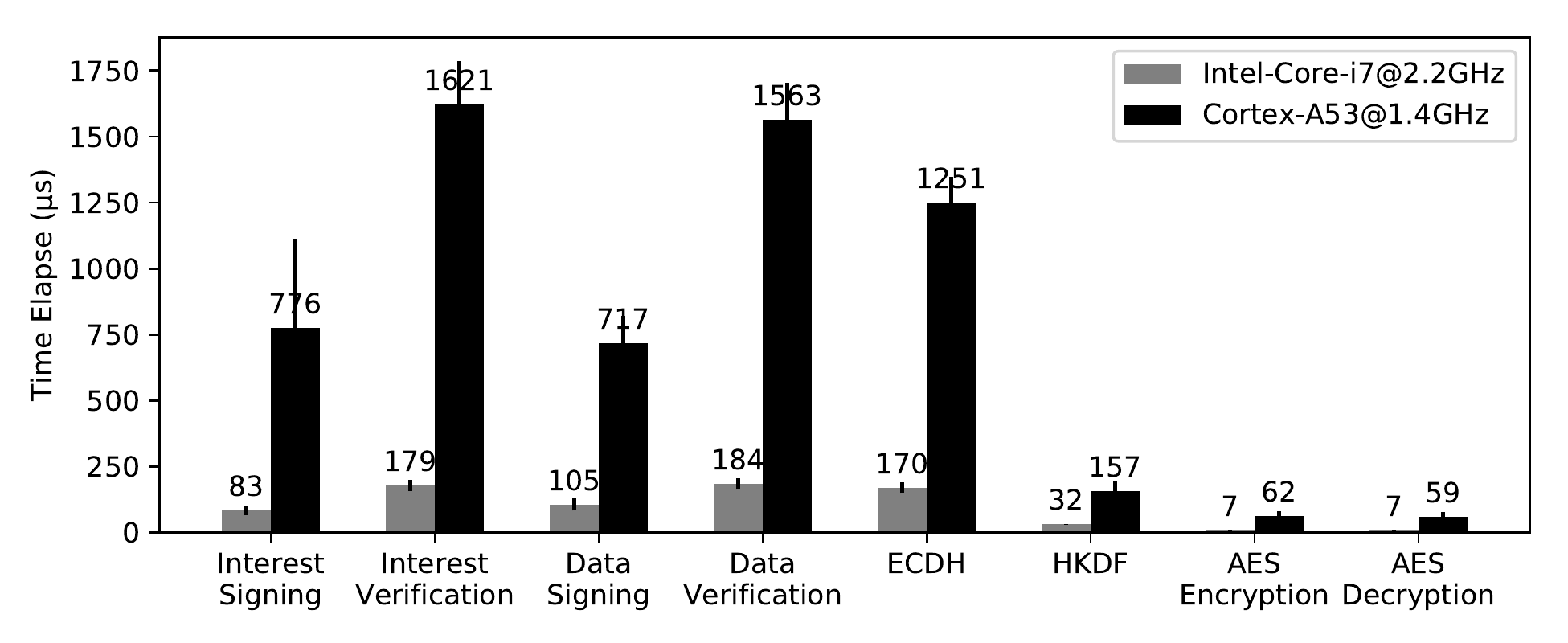}
	\caption{Single Crypto Operation Time Measurement}
	\vspace{-0.3cm}
	\label{fig:crypto-time}
\end{figure}

\para{Formal Analysis of \sysname Protocol}
We formally analyze the latest \sysname protocol with CPSA~\cite{cpsa, cpsa-code}, which generates the shapes of all possible executions to compromise the tested protocol.
In modeling the \sysname protocol into the CPSA, we generalize an identity verification challenge to be the private information sharing between the requester and issuer.
Specifically, the cryptographic protocol of \sysname can be modeled as shown in Table~\ref{table:model} and the notations can be found in Table~\ref{table:notations}.
From the result generated by CPSA, we summarize that \sysname protocol offers protection against attacks enumerated by CSPA because all the possible executions are secure.

\para{Overhead and Performance}
To provide insight into the communication and computation cost, we also measure the packet size and different cryptographic operation time with the C++ \sysname implementation.
In the evaluation, we use \name{/ndn} as the issuer and the requested certificate is under the identity name \name{/ndn/alice}.
Both the issuer's key and requester's key are ECDSA keys over the prime256v1 elliptic curve.
The encryption scheme used in our evaluation is AES-GCM-128.
The evaluation platform is \first a laptop with 8GB memory and 2.2GHz Quad-Core Intel Core i7 and \second a Raspberry Pi 3 Model B+ with 1GB memory and 1.4GHz Cortex A53.

As indicated by the packet size results in Figure~\ref{fig:pkt-size}, \sysname does not pose high bandwidth overhead to the network.
Figure~\ref{fig:crypto-time} shows the time elapse of main cryptographic operations in the \sysname implementation, including Interest and Data packet signature generation and verification operations, ECDH, HKDF~\cite{hkdf}, and AES GCM~\cite{gcm} encryption and decryption.
Among these cryptographic operations, the most time-consuming operation is signature verification, which only takes less than 2ms on a low-power device and is acceptable for practical use.

\subsection{Trial Deployment}
\label{sec:deployment}

We have deployed an old version of \sysname on a single site on the NDN testbed~\cite{ndn-testbed} for more than a year.
A wider deployment is still in progress and the testbed root issuer has already been running over \sysname.
In this section, we first briefly introduce the hierarchical structure of certificates on the NDN testbed and explains how \sysname will be deployed to support it.

\begin{figure}[h]
	\centering
	\includegraphics[width=0.4\textwidth]{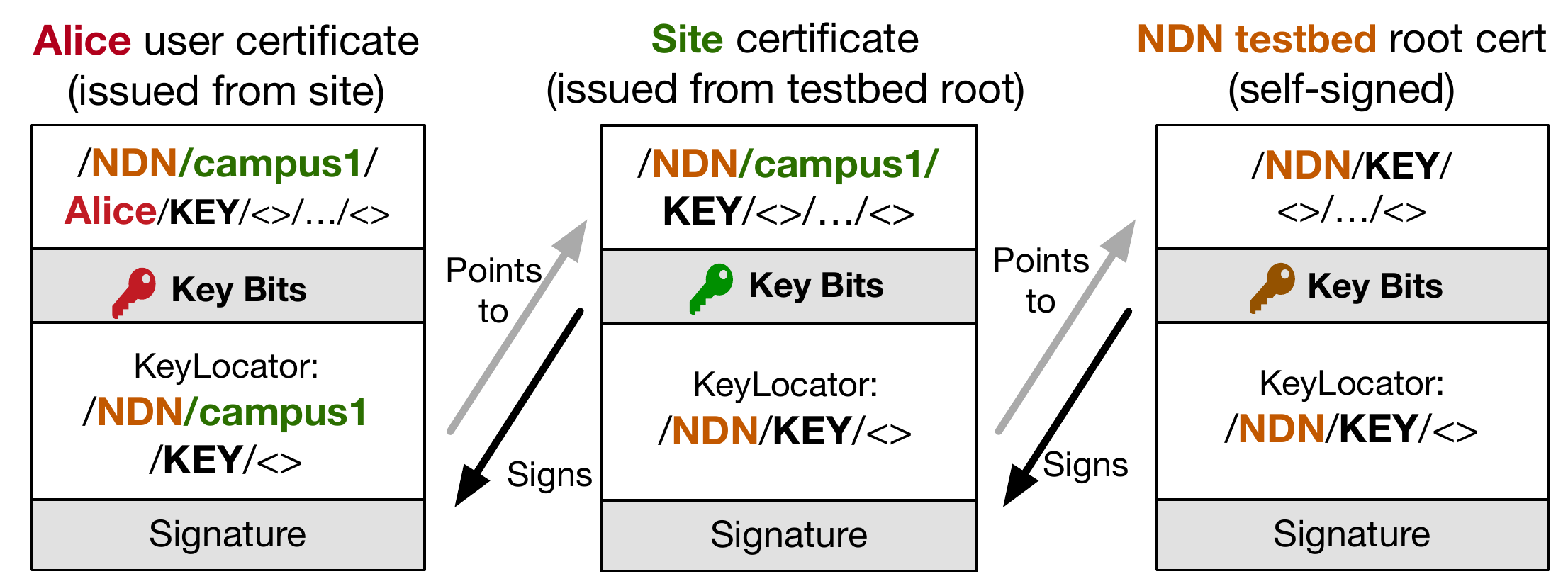}
	\caption{Certificate Chain on the NDN Testbed}
	\label{fig:ndncert-testbed}
\end{figure}

The whole testbed shares one trust anchor (i.e., \name{/ndn}) and each site (e.g., participating institutions or organizations) owns a certificate (e.g. \name{/ndn/campus1}) derived from the trust anchor certificate.
Then each site uses the per-site key to further issue certificates to users from that site.
Testbed users will be configured to trust either the testbed trust anchor certificate or the certificate of the local site.
An example certificate chain from user certificate to testbed anchor certificate is shown in Figure~\ref{fig:ndncert-testbed}.

Therefore, to deploy \sysname, operators from different sites can first request a certificate from the testbed root issuer, where the identity verification challenge can be realized by out-of-band communications (e.g., email exchange or phone call).
After that, each site can use the certified key to run the issuer for users.
Specifically, the site operator can figure out their own form of collaboration with the local name authority and configure their own means to verify requester identity.
To improve usability, users can still be configured to send a certificate request to the testbed issuer and the issuer further forwards users to different sites.

%
%

%% file: sections/challenge-discussion.tex
\section{Discussion}
\label{sec:discussion}


%
%
%

\subsection{Lessons Learned from \sysname Trial Deployment}
\label{sec:discussion:lessons}
\sysname has been under active research and development since 2017 and its design has changed in significant ways over time. We present some of the major design changes and the lessons learned.

\begin{itemize} [leftmargin=*]
\item \textbf{Name assignment:}
In the previous design of \sysname, a certificate issuer served as a name authority as well, assigning a name to a requester when it issues a certificate.
However, throughout the trial deployment, we learned that name assignment in general requires out-of-band knowledge.
Our current design functionally separates name authority from certificate issuer as explained in \S\ref{sec:clarifications}.

\item \textbf{Payload encryption:}
Because only public information, such as a public key, a public-key certificate, and a random application ID, is exchanged in \sysname communication, the previous design of \sysname left all packet exchanges in plain text.
However, as new identity verification challenges were added into the protocol,
and possible attack scenarios were identified by the NDN team, we updated our design to encrypt the payload by default.
\end{itemize}

Additionally, trial usage of \sysname also triggered discussions regarding the content of the KeyLocator field in a packet as we explain next (\S\ref{sec:discussion:multiple-certs}).

\subsection{Same Key, Multiple Certificates}
\label{sec:discussion:multiple-certs}
By design NDN allows one public key to have multiple certificates, as reflected in the NDN certificate name~\S\ref{sec:ndn-certificate}: the same issuer can create a new (renewed/revoked) version of the certificate, and different issuers can issue certificates for the same public key.

With the above in mind, in the current implementation, the KeyLocator field in each signed NDN packet identifies only the key name,
leaving to the consumer to decide which certificate to use to validate the signing key.
However, trial deployments of pilot applications exposed issues in signature validation: although the consumer/verifier's trust policies (trust schemas) make the ultimate decision on acceptable signers and should be able properly construct the name to fetch, additional information is needed to fetch a certificate in \emph{default} cases.
Specifically, when the consumer's trust schema does not include a certificate selection strategy, 
the KeyLocator needs to identify a specific certificate chosen by the producer, so that the consumer can use it for the signature validation.
If the schema specifies the strategy, then it can simply override the certificate instance chosen by the producer.

Although this revelation does not directly relate to the \sysname design, we share this newly gained understanding here with whoever may be working on NDN data verification solutions.

\subsection{Working with Name Authority for Name Assignment}
\label{sec:discussion:name-assignment}

To obtain an NDN certificate, a requester and an issuer must agree on the name of the certificate. This step requires some collaboration between the issuer and the name authority to determine the right name for the certificate.
\begin{figure}[h]
	\centering
	\includegraphics[width=0.25\textwidth]{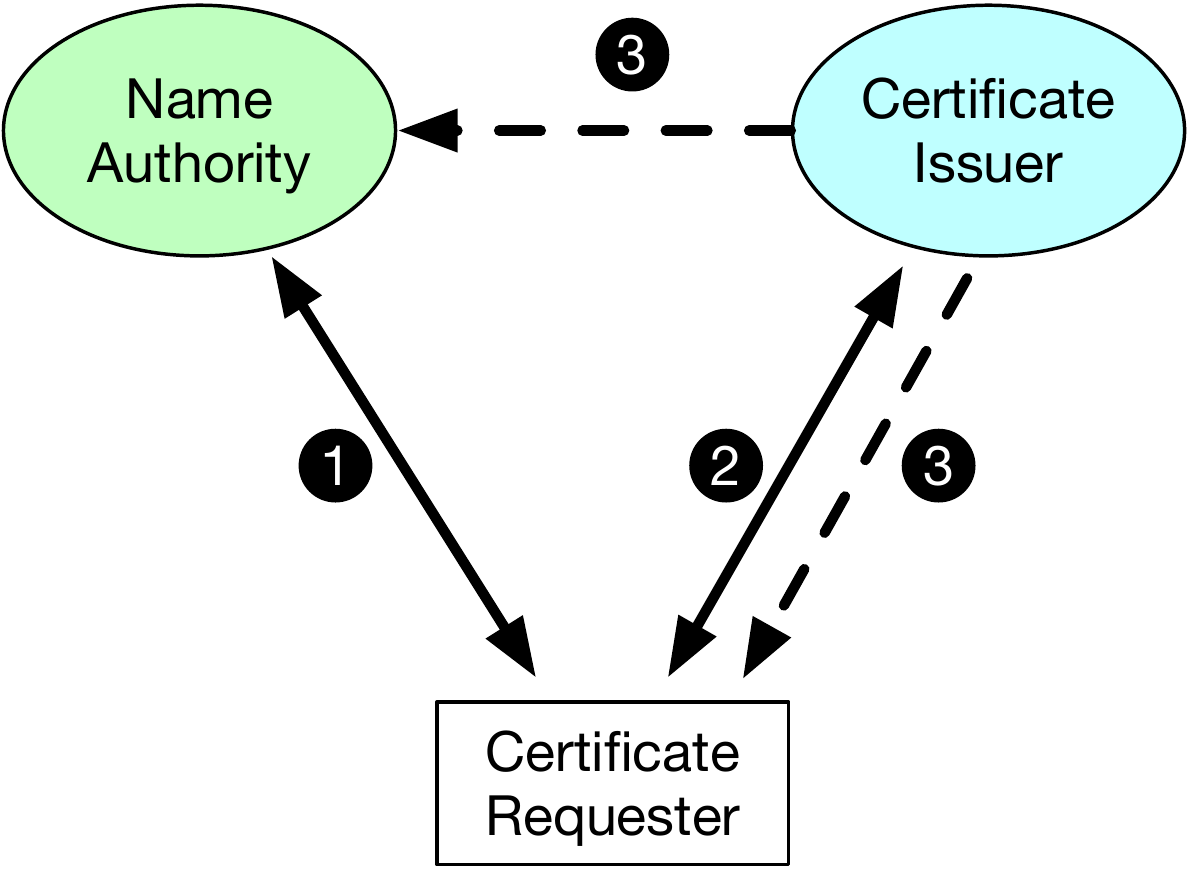}
	\caption{Example collaborations between the certificate issuer and the name authority}
	\label{fig:name-cert}
\end{figure}
This collaboration can be achieved through different means, a typical pattern is shown in Figure~\ref{fig:name-cert}.
A requester $R$ first gets a name from the name authority through an out-of-band channel with name authority's verification of $R$'s identity (\ding{202}).
After that, the name authority gives $R$ an identity assertion (e.g. short-lived secret code) that needs to be presented to the issuer later.
$R$ then runs \sysname client program and sends a certificate request with the identity assertion to the issuer (\ding{203}).
The issuer can verify the assertion through direct or indirect collaboration with the name authority (\ding{204}).

\subsection{Installation of the Trust Anchor}
\label{sec:discussion:anchor}

In order to ensure a certificate issuer is trustworthy, the trust anchor certificate must be pre-installed into each entity via security bootstrapping.
Different application and network scenarios can have different ways to obtain a trust anchor,
however we argue that an out-of-band channel or in-bind channel with out-of-band information is necessary to secure the correctness of the trust anchor information.
\begin{itemize} [leftmargin=*]
\item \textbf{Out-of-band channel}:
The trust anchor certificate can be installed offline.
For example, the trust anchor can be installed with the software or a local network administrator can manually install a trust anchor for each entity.

\item\textbf{In-band channel with out-of-band information}:
A in-band channel can be established after a verification process that utilizing the out-of-band information.
For example, in the case of a college, students may obtain a trust anchor certificate online after verifying the information is truly from the college, e.g., by comparing the hash value of the certificate with the value printed on a hard copy.
\end{itemize}

In IoT scenarios, the installation of the trust anchor can be bundled with device's request of name and certificate when we discuss more about certificate management in IoT next (\S\ref{sec:discussion:iot}).

\subsection{NDN Certificate and Name Management in Internet of Things}
\label{sec:discussion:iot}

While the \sysname design considers generic use cases, in Internet of Things (IoT) scenarios, the installation of trust anchor, the name assignment, and the process of certificate issuance can be bundled together for simplicity and efficiency~\cite{ssp, evolvingzhang, compagno2016onboardicng}.

In such a process, based on certain out-of-band information (e.g., QR code scanning), a mutual authentication can be established between an IoT device with the controller who acts as the trust anchor, the name authority proxy, and certificate issuer.
After that, the device can install the local trust anchor certificate and the controller can assign the name with the help from the higher layer (e.g., human) and issue the certificate for the device.



\subsection{Grandchild Namespace Certificates}
\label{sec:discussion:grandchild}


When a namespace owner assigns a sub-namespace to another entity and issues it a certificate, the issued certificate is a proof of the namespace ownership delegation from the issuer to the requester.
As we discussed in \S\ref{sec:issue-cert}, NDN allows such operation to occur recursively.
This means that an entity needs to set up its own certificate issuer program in order to be able to issue sub namespace certificates.

However, an entity may not be able or willing to directly manage the delegated namespace, either because of the limited resources (e.g., in IoT cases), the need to stay available for the requesters, or simply because of the lack of experience to setup the software.
In these cases, we may allow ``(great-*)grandparents'' to issue certificates to the sub-delegated namespace identities, \emph{but only if this is permitted by the trust schema rules}.
In other words, when the schema does not allow such delegation, even though the certificate can be physically issued, it will be discarded during the rule validation stage.

%
%

%% file: sections/conclusion.tex
\section{Conclusion}


To support the ubiquitous use of public key cryptography to secure data in NDN,
this paper presented the design of \sysname to provide usable certificate management.
The design of \sysname is based on our latest understanding of several key concepts and the relations among them, including
name assignment versus certificate management,
authentication versus trust, and
certificate issuers versus trust anchors.

Clarification of these concepts helps us further appreciate the fundamental differences between today's deployed TLS certificate management system and 
NDN certificate management.
The latter is established on the foundation of 
\first~naming data by using the same structured, semantically meaningful names throughout the protocol stack, 
\second~clear separation between authentication and trust, and
\third~flexible certificate issuance ability to support least privilege principles.
As a result, NDN certificate management avoids all the issues, as identified in
\S\ref{sec:motivation}, in today's deployed solutions. 

In addition, we recognize that NDN/ICN can facilitate the certificate management by providing simple certificate retrieval and high certificate availability.
As future work, we will expand the deployment of \sysname from the NDN testbed into all NDN-enabled devices and applications; such real trials will help deepen our understanding of how data-centric security should be developed.

